\documentclass[rnote]{aa}  
\usepackage{graphicx}
\usepackage{color}
\usepackage{epsfig}
\usepackage{supertabular}
\usepackage{txfonts}
\begin{document}
   \title{A search for massive UCDs in the Centaurus Galaxy Cluster \thanks{Based on observations obtained in service mode at the VLT (programme 080.B-0307)}}

   \author{Steffen Mieske
          \inst{1}
          \and
          Michael Hilker\inst{2}\and Ingo Misgeld\inst{2,3}\and Andr\'{e}s Jord\'{a}n\inst{4,5} \and Leopoldo Infante\inst{4} \and Markus Kissler-Patig\inst{2}
         }

   \offprints{S. Mieske}

   \institute{European Southern Observatory, Alonso de Cordova 3107, Vitacura, Santiago, Chile \and
European Southern Observatory, Karl-Schwarzschild-Strasse 2, 85748 Garching bei M\"unchen, Germany\and
Argelander Institut f\"ur Astronomie, Auf dem H\"ugel 71, 53121 Bonn, Germany\and
     Departamento de Astronom\'{\i}a y Astrof\'{\i}sica, Pontificia
Universidad Cat\'olica de Chile, Casilla 306, Santiago 22, Chile 
\and Harvard-Smithsonian Center for Astrophysics, 
        60 Garden St., Cambridge, MA 02138
}

   \date{}

 
  \abstract 
{We recently initiated a
search for ultra-compact dwarf galaxies (UCDs) in the Centaurus
galaxy cluster (Mieske et al. 2007), resulting in the discovery of 27 compact
objects with $-12.2<M_V<-10.9$ mag. Our overall survey completeness
was 15-20\% within $120$ kpc projected clustercentric distance.} 
{In order to better constrain the luminosity
  distribution of the brightest UCDs in Centaurus, we continue our search by substantially improving
  our survey completeness specifically in the regime $M_V<-12$ mag
  ($V_0<22.3$ mag). }  
{Using VIMOS at the VLT, we obtain low-resolution spectra of 400
  compact objects with $19.3<V_0<21.3$ mag ($-14<M_V<-12$ mag
  at the Centaurus distance) in the central 25$'$ of the Centaurus
  cluster, which corresponds to a projected radius of $\sim$150 kpc. Our survey yields complete area coverage within $\sim$$120$ kpc. }
{For 94\% of the sources included in the masks we successfully measure
  a redshift. Due to incompleteness in the slit assignment, our final
  completeness in the area surveyed is 52\%.  Among our targets we
  find three new UCDs in the magnitude range $-12.2<M_V<-12$ mag,
  hence at the faint limit of our survey. One of them is covered by
  archival HST WFPC2 imaging, yielding a size estimate of
  $r_h\lesssim$ 8-9 pc. At 95\% confidence we can reject the
  hypothesis that in the area surveyed there are more than 2 massive
  UCDs with $M_V<-12.2$ mag and $r_{\rm eff} \lesssim$70 pc. Our
  survey hence confirms the extreme rareness of massive UCDs. We find
  that the radial distributions of Centaurus and Fornax UCDs with
  respect to their host clusters' centers agree within the 2$\sigma$
  level.}
{}

\titlerunning{Massive UCDs in Centaurus}


   \maketitle 
%

\section{Introduction}
A new class of compact stellar systems called 'ultra-compact dwarf
  galaxies' (UCDs; Phillipps et al.~\cite{Philli01}) has been
  established during the last decade (Hilker et al.~\cite{Hilker99};
  Drinkwater et al.~\cite{Drinkw00},~\cite{Drinkw03}; Hasegan et
  al.~\cite{Hasega05}; Jones et al.~\cite{Jones06}; Mieske et
  al.~\cite{Mieske07}; Firth et al.~\cite{Firth07}; Misgeld et
  al.~\cite{Misgel08}). UCDs are characterised by typical luminosities
  of $-13.5<M_V<-11.0$ mag, half-light radii of $10<r_h<100$ pc and
  masses of $2\times 10^6 < m < 10^8\ \mathrm{M_{\odot}}$. An
  intriguing finding from recent studies is that on average, the
  dynamical M/L ratios of UCDs are about twice as large as those of
  Galactic globular clusters of comparable metallicity (e.g.  Hasegan
  et al.~\cite{Hasega05}; Hilker et al.~\cite{Hilker07}; Evstigneeva
  et al.~\cite{Evstig07}; Rejkuba et al.~\cite{Rejkub07}; Mieske et
  al.~\cite{Mieske08b}).  Indications exist that M/L ratios may
  be somewhat higher for UCDs in Virgo than in Fornax (e.g.  Hasegan
  et al.~\cite{Hasega05}; Hilker et al.~\cite{Hilker07}; Evstigneeva
  et al.~\cite{Evstig07}), which could be explained by differences in
  age, stellar mass function, or, dark matter content.

In our efforts to broaden the environmental baseline of UCD research,
we have embarked on UCD searches in the Centaurus galaxy cluster
(Mieske et al.~\cite{Mieske07}) and the Hydra I galaxy cluster
(Misgeld et al.~\cite{Misgel08}), based on data obtained with VIMOS at
the VLT (program 076.B-0293). From the UCD search in Centaurus, we
found 27 compact objects with radial velocities consistent with them
being members of Centaurus, covering an absolute magnitude range $-$12.2
$<$ M$_V$ $<$ $-$10.9 mag. Their distribution in magnitude and space was
found to be consistent with that of the GC population. We did not find
very luminous UCDs with $-13.5 < M_V< -12.2$ mag as found in the
Virgo and Fornax cluster (Jones et al.~\cite{Jones06}, Chilingarian \&
Mamon~\cite{Chilin08}, Drinkwater et al.~\cite{Drinkw00}), which may
be due to the moderate overall completeness of 15-20\% within
120 kpc: the survey covered about 50-60\% of the total area within 120 kpc,
within which only 30\% of photometrically selected sources were
observed, due to too high candidate density (see Mieske et
al.~\cite{Mieske07} and Fig.~\ref{map}).

In this Research Note, we report on the second part of our search for
UCDs in Centaurus with VIMOS (program 380.B-0207), with the aim to
better constrain their luminosity distribution at the bright
end. We increase the area coverage and focus on the bright luminosity
regime $M_V<-12$ mag. Our medium-term strategy is to derive dynamical
mass estimates for UCDs beyond the Fornax and Virgo clusters to
investigate in depth whether average M/L ratios of UCDs systematically
vary with environment. Technically, ground-based medium-to-high
resolution spectroscopy (R$\sim$10000) is only possible for objects
with $V\lesssim$21.5 mag (e.g. Mieske et al.~\cite{Mieske08b}, Chilingarian
et al.~\cite{Chilin08}), which translates to a feasibility limit of
$M_{V,0} \lesssim -12$ mag at the distance of the Centaurus cluster
((m-M)$\simeq$33.3 mag, Mieske et al.~\cite{Mieske05}). This is
another driver for focusing on the bright luminosity regime $M_V<-12$
mag.

\section{The data}
The data for this publication were obtained in service mode with the
VIsible MultiObject Spectrograph VIMOS (Le Fevre et
al.~\cite{Lefevr03}) mounted on UT3 Melipal at the VLT (programme
380.B-0207). VIMOS allows simultaneous observing of 4 quadrants, each
of dimension $7'\times 8'$, and separated by about 2$'$. We observed
four multi-object spectroscopy (MOS) pointings close to NGC 4696, the
central galaxy of the main cluster component Cen30 (see
Fig.~\ref{map}). Two of those pointings had been targeted already as
part of our previous observing program in Period 76 (Mieske et
al.~\cite{Mieske07}).  However, given the slit allocation completeness
of about 30\% in that run, we re-observed these pointings. We observed
two further pointings slightly offset, with the aim to fill the chip
gaps, increasing the area coverage within $\sim$120 kpc to almost
100\% (see Figs.~\ref{map} and~\ref{map2}). Within the four pointings,
a slit could be allocated for 54\% of the photometrically selected objects
(see Sect.~\ref{spectroscopic}).

\subsection{Candidate selection}
\label{candsel}
The candidates for our search for bright UCDs were selected from the
VIMOS pre-imaging in the $V$ and $R$ filters which were taken under clear conditions. Prior to applying any selection, we matched the
detections in V,R to the catalog of well calibrated FORS photometry
(Mieske et al.~\cite{Mieske05}) of the central Centaurus cluster in V and I,
whose areas overlap with the VIMOS pre-imaging. From this matching we
were able to verify that the V-band VIMOS zeropoints
available from the ESO QC web pages\footnote{http://www.eso.org/observing/dfo/quality/index\_vimos.html} for the date of the pre-imaging (17-01-2008)
were accurate to within 0.03-0.05 mag.

For de-reddening the apparent magnitudes we used Schlegel et
al.~(\cite{Schleg98}).  To select sources as compact object
candidates, we defined three criteria regarding size, colour and
luminosity. 

\begin{figure}[h!]
\begin{center}
  \epsfig{figure=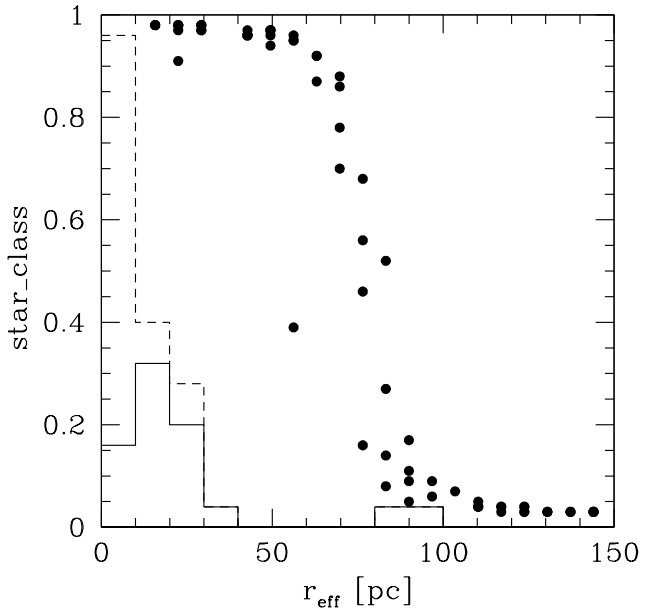,width=8.6cm}
  \caption{This plot illustrates to which upper limit of UCD effective
  radius our candidate selection criterion for unresolved sources
  corresponds to. The y-axis shows the SExtractor star-classifier
  value (Bertin \& Arnouts~\cite{Bertin96}), which is designed to be 1
  for an unresolved source (``star''), and 0 for a resolved source
  (``galaxy''). We have simulated seeing convolved UCD images on top
  of our VIMOS pre-images, based on the structural parameters of UCD3
  (Hilker et al.~\cite{Hilker07}), and assuming a Centaurus cluster
  distance of 45 Mpc (Mieske et al.~\cite{Mieske05}). The seeing was
  0.85$''$, corresponding to $\simeq$ 190 pc at the Centaurus
  distance. UCD3 itself has $r_{\rm eff}\sim 90$pc. To simulate UCDs
  of a range of sizes, we scaled the surface brightness profile of
  UCD3 to smaller / larger radii. The plot shows that the SExtractor
  star classifier value flips from ``star'' to ``galaxy'' for $r_{\rm
  eff} \gtrsim$70 pc. The solid histogram is the size distribution of
  known UCDs with $M_V<-12$ mag, arbitrarily normalised. The dashed
  histogram is the size distribution of all UCDs (hence $M_V<-11$
  mag), normalised by the same factor as the dashed histogram.}
\label{size}
\end{center}
\end{figure}

1. Be unresolved on the VIMOS pre-imaging (as judged by SExtractor
star-galaxy separator, Bertin \& Arnouts~\cite{Bertin96}). At the
distance of the Centaurus cluster (45 Mpc, Mieske et
al.~\cite{Mieske05}), the typical PSF FWHM of 0.85$''$ corresponds to
$\approx$ 190pc. In Fig.~\ref{size} we show that the limit up to which
SExtractor classifies a source as unresolved corresponds to $r_{\rm
  eff}\simeq$ 70 pc at the Centaurus cluster distance. Our size
selection criterion hence encompasses all known UCDs except the two
most massive ones, each of which have $r_{\rm eff}\simeq$ 100 pc
(Evstigneeva et al. 2008). This corresponds to $\sim$ 95\% of all
known UCDs, and $\simeq$ 85\% of known UCDs with $M_V<-12$
mag. \footnote{A complementary observing campaign targeting canonical
  dwarf galaxies -- including resolved UCD candidates -- in Centaurus
  has been approved for ESO observing period P83.}

2. Have de-reddened colours $0.42<(V-R)_0<0.9$ mag. This (V-R)$_0$
range corresponds to a (V-I)$_0$ range of 0.65 to 1.50 mag (see
Fig.~\ref{map2}), which is the colour range typically covered by GCs
(e.g. Gebhardt \& Kissler-Patig~\cite{Gebhar99}, Larsen et
al.~\cite{Larsen01}, Kundu \& Whitmore~\cite{Kundu01}).  This also
covers the colour range of UCDs discovered in the first part of our
search (Mieske et al.~\cite{Mieske07}). Fig.~\ref{map2} shows the
directly measured (V-R)$_0$ colours, and the corresponding (V-I)$_0$
scale. This scale is derived from matching the V,R pre-imaging
photometry of unresolved sources with V,I photometry from spatially
overlapping FORS data (see above; Mieske et al.~\cite{Mieske05}). The
derived scaling in apparent magnitude space is

\begin{equation}
(V-I)=-0.163+1.911\times(V-R)
\label{vical}
\end{equation}

with an rms of 0.11 mag.

3. Have de-reddened apparent magnitudes $19.3<V_0<21.3$ mag
($-14<M_V<-12$ mag). The faint magnitude cut is more than a
magnitude brighter than in the first part of our search (Mieske et
al.~\cite{Mieske07}).

\subsection{Spectroscopic observations}
\label{spectroscopic}
Within our 16 masks (4 pointings $\times$ 4 quadrants) the VIMOS mask
creation software VMMPS enabled the allocation of slits for 412
objects (minimum slit length 6$''$), compared to a total of 766
photometrically selected sources.  We were able to measure redshifts
for 389 out of those 412 sources.  Our completeness in the entire area
surveyed is hence 51\% (Fig.~\ref{completeness}), about three times
higher than in the previous part of our survey (Mieske et
al.~\cite{Mieske07}). Within the central 120 kpc, the completeness is
52\%, only marginally higher.

We used the medium resolution MR grism with the order sorting filter
GG475. This covers the wavelength range from 4800 to 10000 {\AA} at a
dispersion of 2.5 {\AA} per pixel. The average seeing for the
spectroscopic observations was around 0.8$''$, at a slit width of
1.0$''$.  With a pixel scale of 0.2$''$, the instrumental resolution
(FWHM) is 10-12 \AA, corresponding to a velocity resolution of
$\sim$600 km/s.  For each pointing the total exposure time was 2100
seconds. Arc-lamp exposures for wavelength calibration were attached
to each science exposure.

\subsection{Data reduction}
\label{reduction}
\begin{figure*}[h!]
\begin{center}
  \epsfig{figure=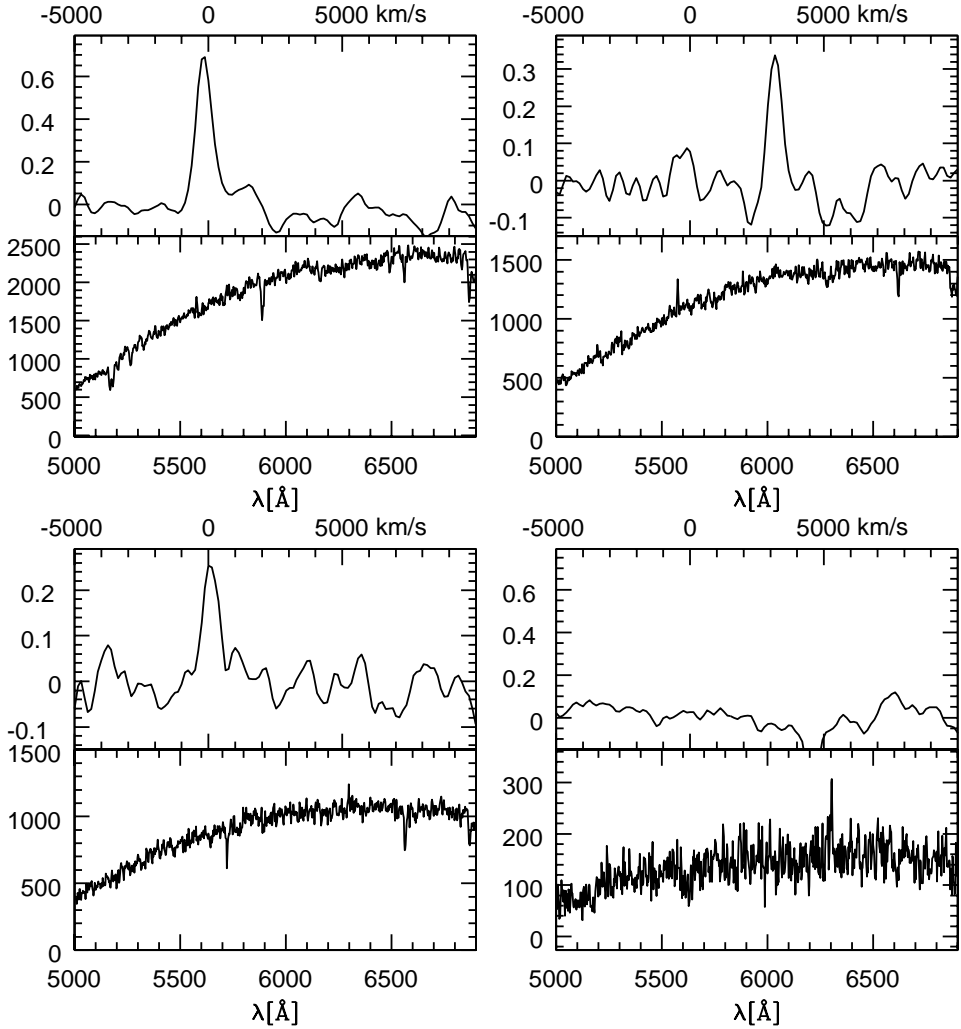,width=14cm}
  \caption{Four example spectra of our VIMOS data-set, with the corresponding cross-correlation results for radial velocity measurement indicated on top. Y-axis units are flux in ADU for the bottom plots, and cross-correlation height $h$ for the top plots. Objects from upper left to bottom right: A foreground star with cross-correlation confidence level R$>5.5$;  one of the three Centaurus cluster members (CCOS J192.129-41.401), also with $R>5.5$; a foreground star with $R=5.0$, whose radial velocity measurement was accepted; a source for which no radial velocity could be measured.}
\label{spectra}
\end{center}
\end{figure*}

For the data reduction from 2D raw spectra to wavelength calibrated 1D
spectra we used the recipe vmmosobsstare provided by the ESO VIMOS
pipeline\footnote{http://www.eso.org/projects/dfs/dfs-shared/web/vimos/vimos-pipe-recipes.html}.
This recipe performs bias subtraction, flat field division, wavelength
calibration, and spectrum extraction. Fig.~\ref{spectra} shows four
examples of calibrated 1D spectra.

\begin{figure*}[h!]
\begin{center}
  \epsfig{figure=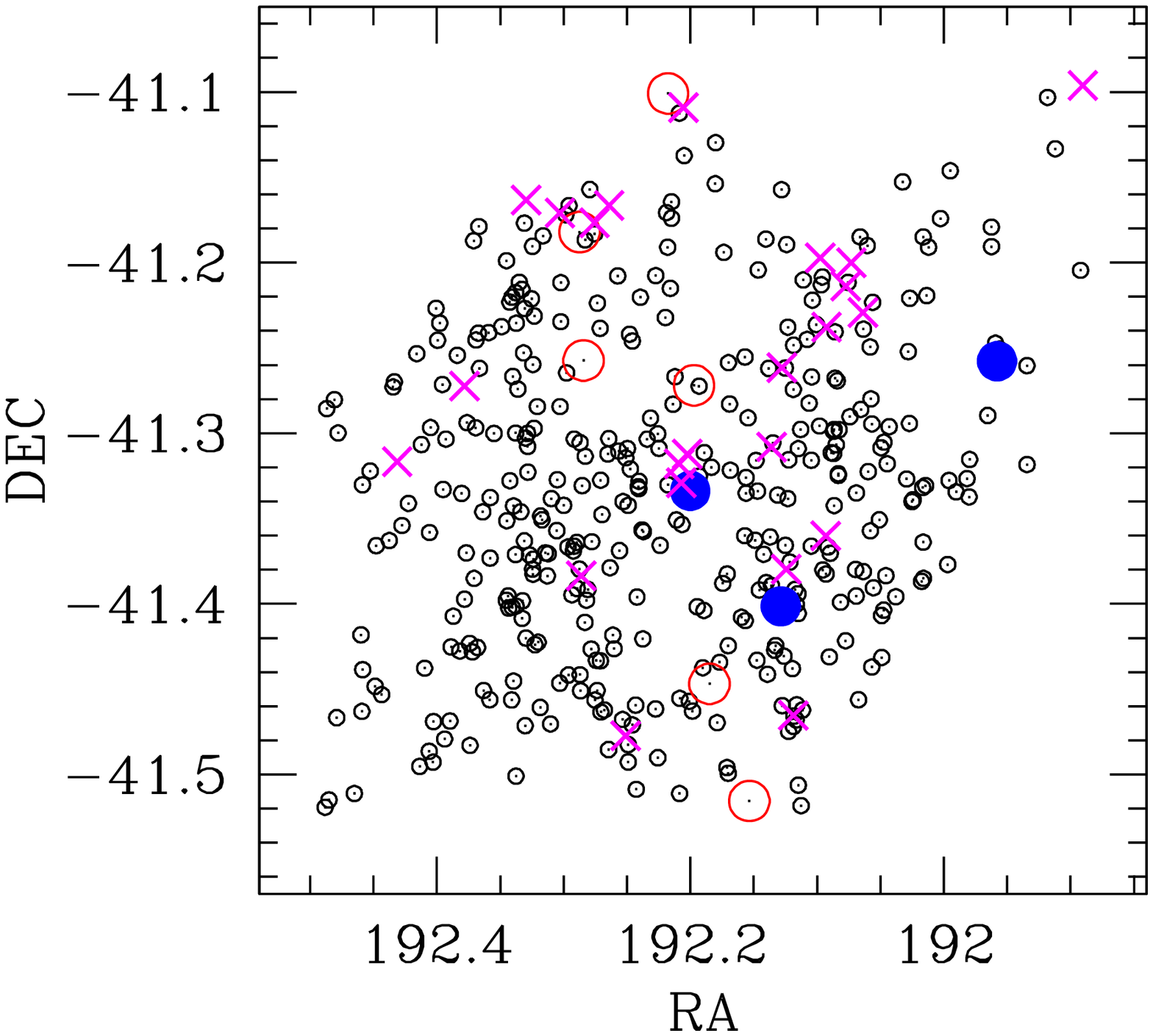,width=8.6cm}
  \epsfig{figure=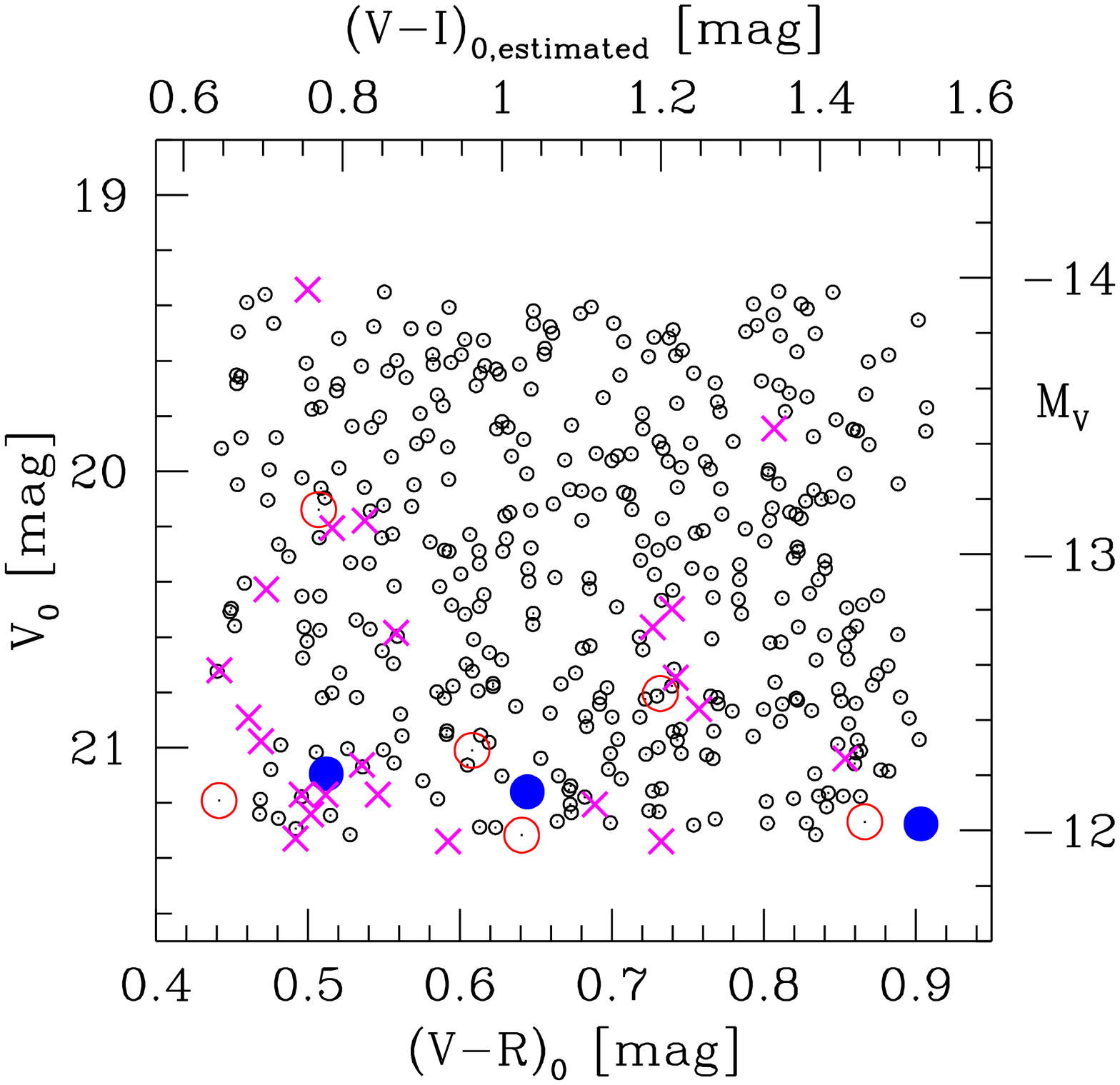,width=8.6cm}
  \caption{Map (left) and CMD (right) of the objects observed in the
  VIMOS P80 run. Dots indicate all objects to which a slit was
  assigned. Small (black) circles indicate foreground stars. Large
  (red) open circles indicate background sources ($v_{\rm rad}>6000$
  km/s). Magenta crosses mark objects for which no radial velocity
  could be measured. Filled (blue) circles indicate objects with
  radial velocities in the Centaurus cluster range, hence the UCDs
  found. In the CMD, the upper x-axis shows the approximate scale of
  (V-I)$_0$ colours. The scaling of (V-R) with (V-I) was derived from
  matching the V,R pre-imaging photometry with spatially overlapping
  FORS V,I photometry (Mieske et
  al.~\cite{Mieske05}).}
\label{map2}
\end{center}
\end{figure*}

\begin{figure*}[h!]
\begin{center}
  \epsfig{figure=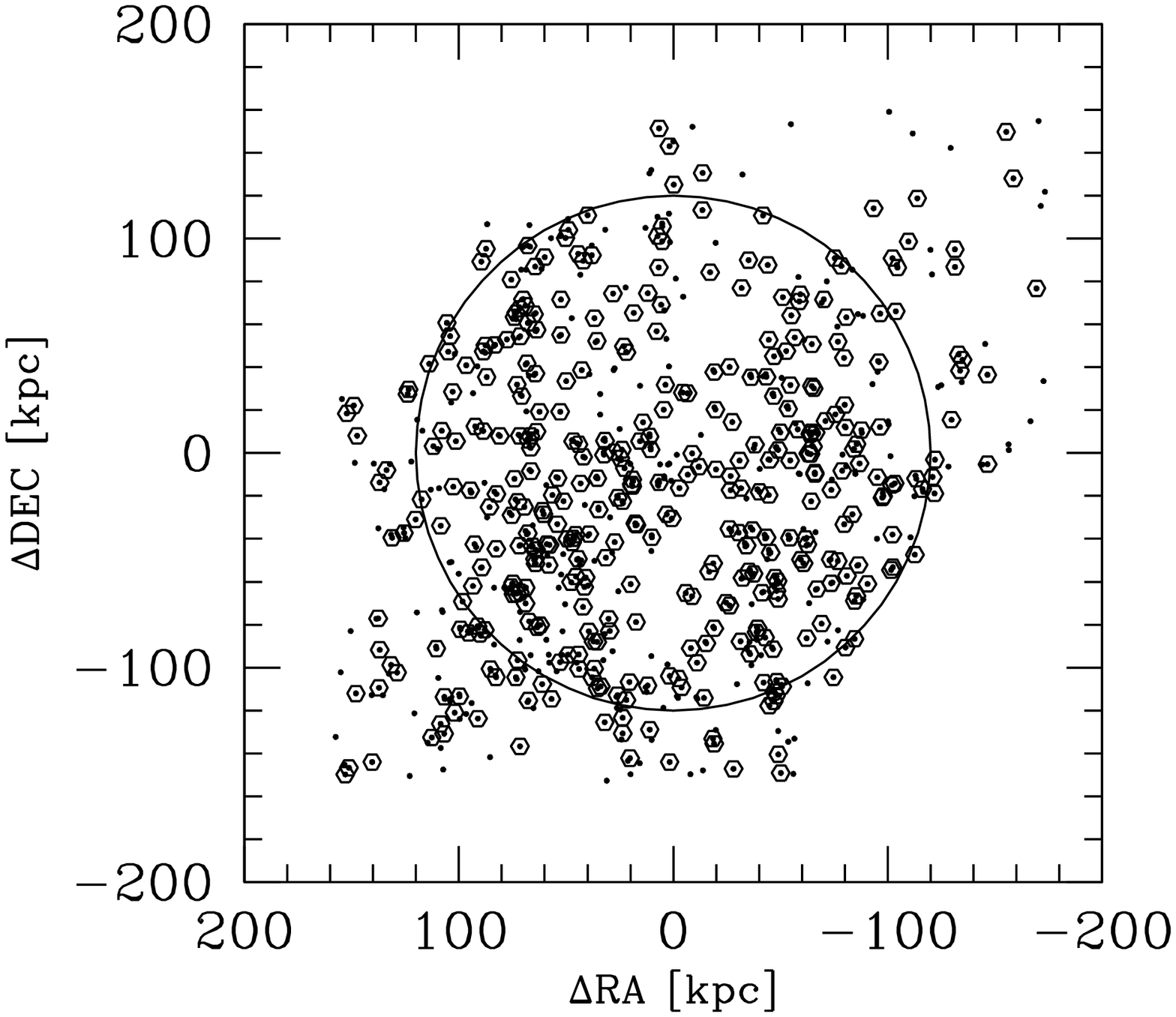,width=8.6cm}
  \epsfig{figure=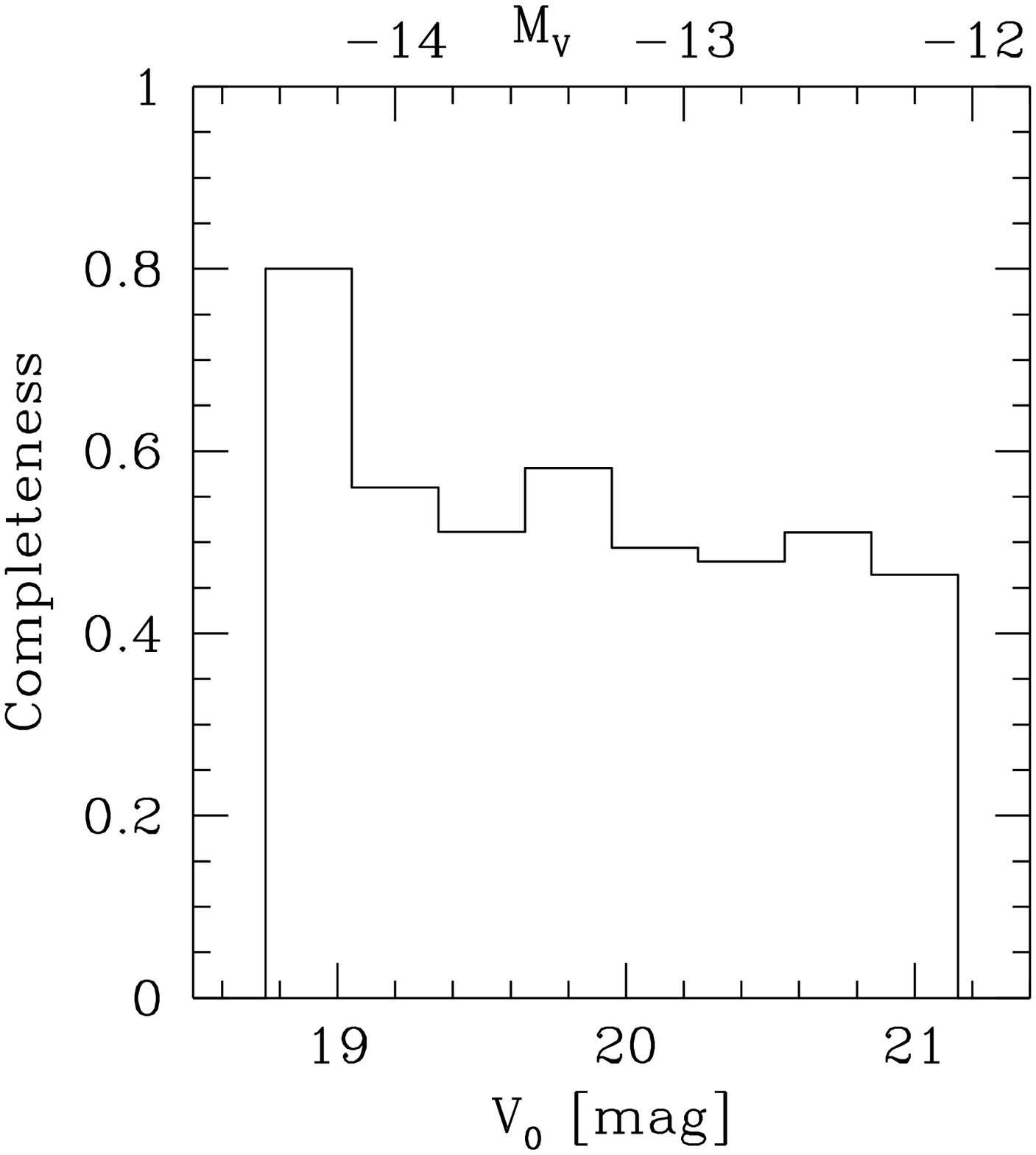,width=8.6cm}
  \caption{Completeness of our survey. {\bf Left panel:} Plotted are
  all photometrically selected objects (small dots) and successfully
  observed objects (small hexagons). Note the difference to
  Fig.~\ref{map2}, where only objects included in the masks are
  plotted. The circle indicates a projected radius of $\sim$120 kpc at
  the Centaurus cluster distance. {\bf Right panel:} Ratio of
  successfully observed objects to all photometrically selected
  objects within the central 120 kpc, as a function of magnitude. }
\label{completeness}
\end{center}
\end{figure*}

The radial velocity measurements of the calibrated 1D spectra were
performed via cross-correlation using the IRAF task fxcor (Tonry \&
Davis~\cite{Tonry79}) in the RV package.  As template for
cross-correlation we used a synthetic spectrum created to resemble a
typical early-type galaxy (Quintana et al.~\cite{Quinta96}). This
template has proven most reliable for such kind of radial velocity
surveys (e.g. Mieske et al.~\cite{Mieske04a}, Misgeld et
al.~\cite{Misgel08}). For a measurement to be accepted as reliable, we
demanded the cross-correlation confidence value $R$ to be larger than
5.5. We then re-run fxcor for those spectra for which $R<5.5$ was
achieved in the first run, and accepted $v_{rad}$ measurements for
those that showed clearly identifiable cross-correlation peaks, see
Fig.~\ref{spectra} for an example.  For more than 90\% of our observed
sources we could reliably measure a redshift (see
Fig.~\ref{map2}). The radial velocity measurement errors were of the
order 50-100 km/s.  As a cluster membership criterion we required
$1750<v_{rad}<5550$ km/s, excluding both foreground stars and
background galaxies.

\begin{figure}[h!]
\begin{center}
  \epsfig{figure=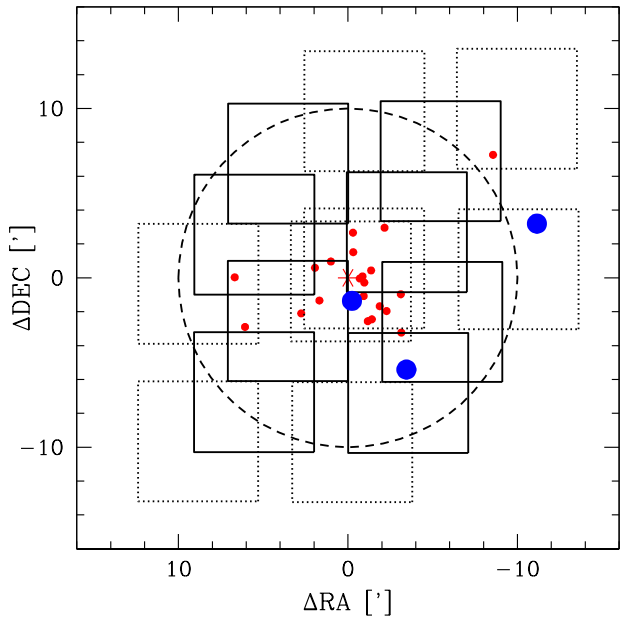,width=8.6cm}
  \caption{Map of the central Centaurus cluster. The relative
  coordinates are with respect to NGC 4696, the central galaxy of the
  Cen30 subcluster. The dotted squares indicate the VIMOS pointing
  observed in P76 (Mieske et al.~\cite{Mieske07}). Note that the VIMOS
  field-of-view consists of four quadrants. The small (red) dots
  indicate the UCDs detected in that survey, covering a magnitude
  range $-12.2<M_V<-10.8$ mag. For the new run in P80 which is
  reported upon in this paper, we re-observed the dotted pointings
  (see text) with two masks per quadrant, added two more pointings
  (solid squares), and focused only on $M_V<-12$ mag. The large (blue)
  dots indicate the three UCDs found in this new run (see
  Sect.~\ref{results}). The dashed circle indicates a projected
  clustercentric distance of $\sim$120 kpc, within which most Fornax
  and Virgo UCDs are found (see Thomas et al.~\cite{Thomas08} for a
  sample of fainter intra-cluster UCDs in Fornax at larger projected
  radii).}
\label{map}
\end{center}
\end{figure}

\begin{figure*}[h!]
\begin{center}
  \epsfig{figure=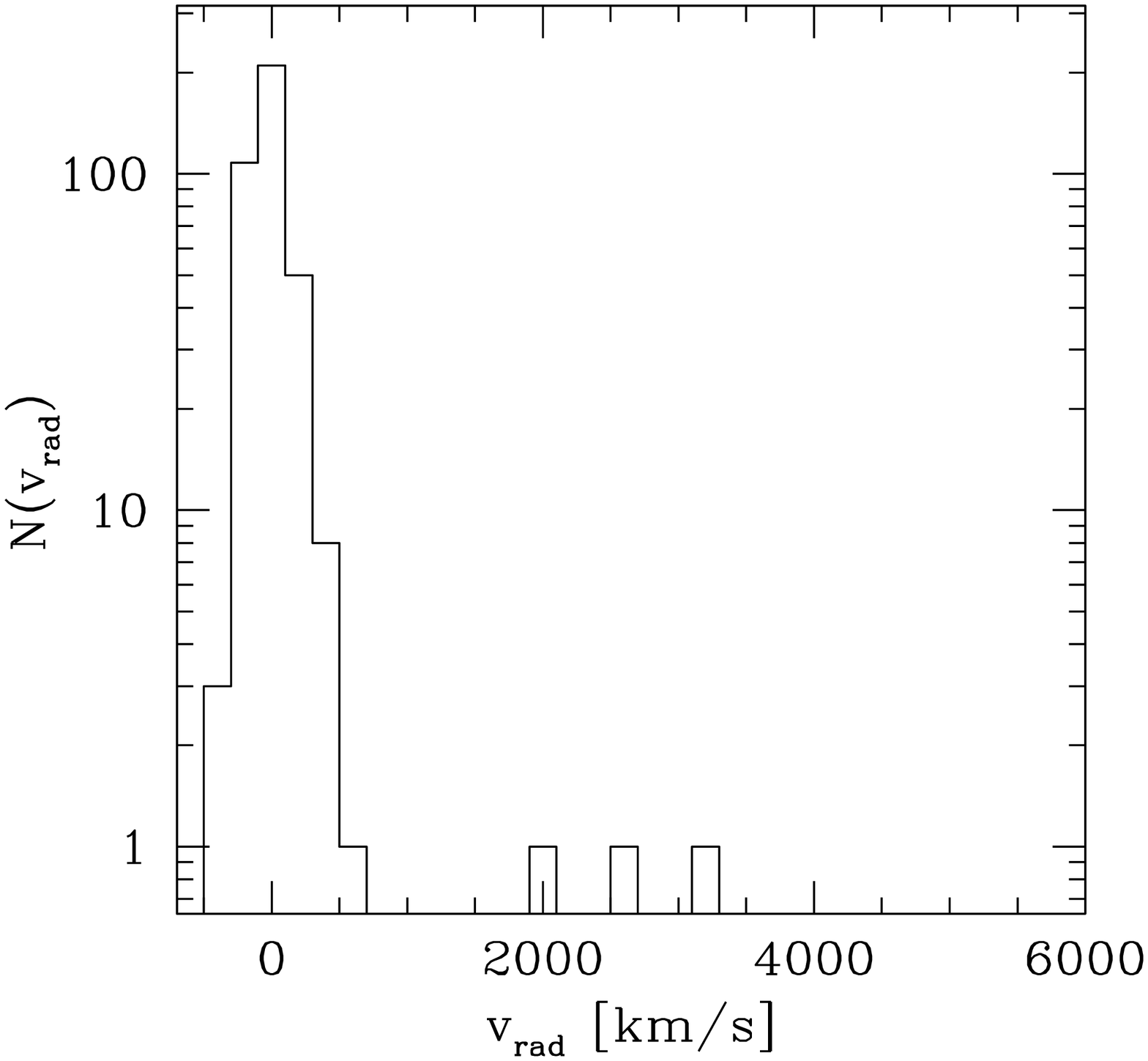,width=6.5cm}
  \epsfig{figure=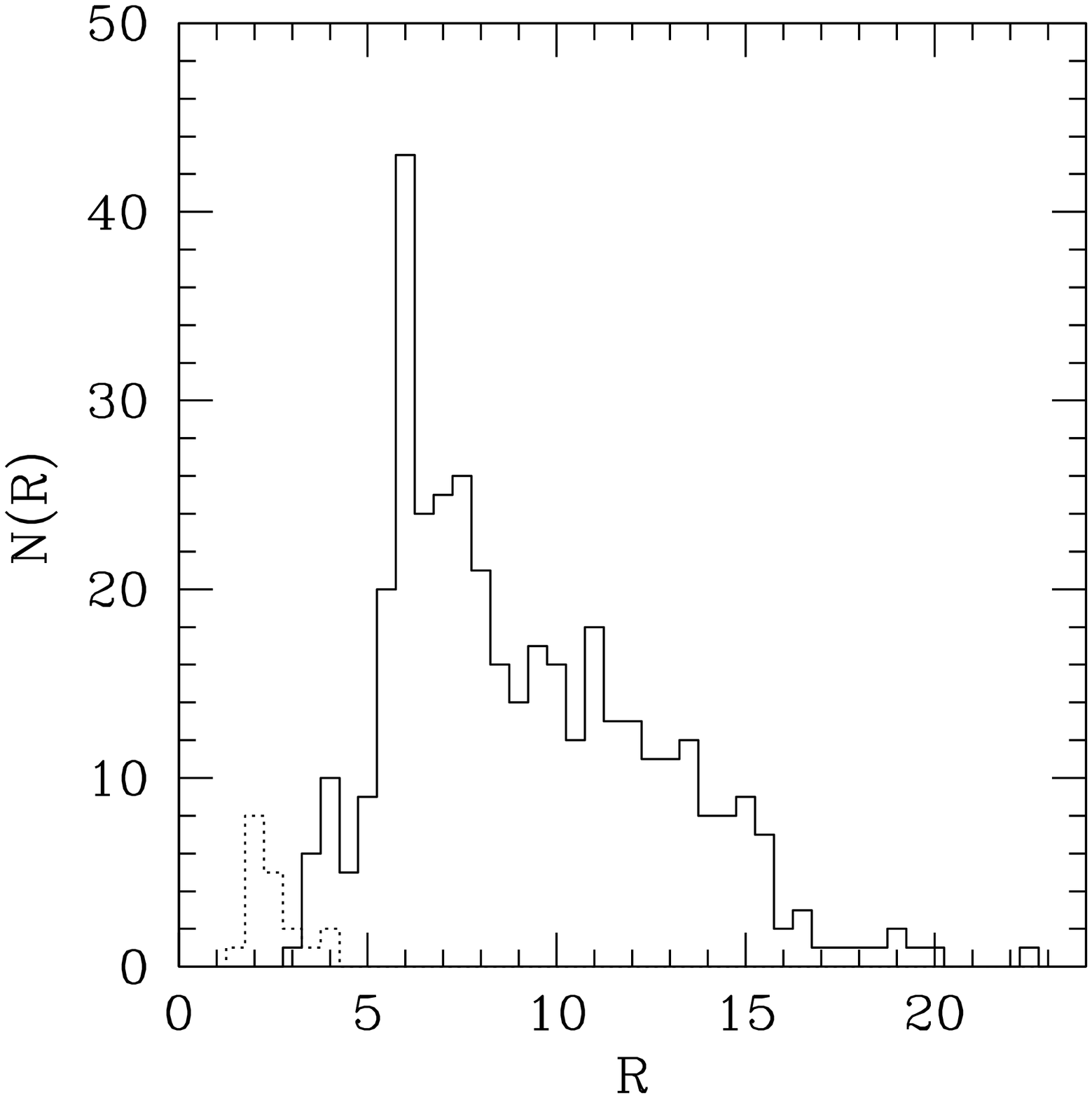,width=6.5cm}
  \caption{{\bf Left panel:} radial velocity histogram of the observed
  sources. The background regime ($>6000$km/s) is excluded. The
  three Centaurus cluster members are clearly distinguishable. {\bf Right
  panel:} histogram of the confidence level $R$ achieved in the radial
  velocity measurement via cross-correlation with a template
  spectrum. The solid histogram indicates the sources with reliable
  measurement, the dotted histogram indicates sources without an
  acceptable cross-correlation result.}
\label{vradhist}
\end{center}
\end{figure*}

\section{Results}
\label{results}
Fig.~\ref{vradhist} shows a radial velocity histogram of the 389
sources with measured redshifts.  Of these 389 sources, 380 are
foreground stars, and six objects are background galaxies with
$9400<v_{rad}<41000$ km/s\footnote{The list of foreground stars and
background galaxies including their coordinates is available upon
request from smieske@eso.org}. Only three objects are members of the
Centaurus cluster. At the cluster's distance modulus ((m-M)=33.3 mag,
Mieske et al. 2005), they cover the magnitude range $-12.2<M_V<-12.0$
(Fig.~\ref{map2}), at the faint limit of our survey. Table~\ref{table}
shows the properties of these three confirmed UCDs. We list their
$V_0$ and $(V-R)_0$ magnitudes, their radial velocities and errors,
and the confidence level $R$ of the radial velocity measurement. Also
given is the estimated $(V-I)_0$ colour, as derived from
equation~\ref{vical}.  For one of the UCDs, CCOS J192.200-41.334,
archival HST imaging in the F555W filter (WFPC2, Proposal 5956, PI
Sparks) is available. To measure its size, we use the program KINGPHOT
(Jord\'an et al.~\cite{Jordan04} and~\cite{Jordan05}), which was
already successfully applied to measure half-light radii $r_h$ of GCs
in Virgo and Fornax, and UCDs in Centaurus (Jord\'an et
al.~\cite{Jordan05}~and~\cite{Jordan07}, Mieske et
al.~\cite{Mieske07}). From this fit, we derive a 2$\sigma$ upper limit
of the projected half-light radius of 0.43 WFPC2 wide-field pixel,
corresponding to $r_h\lesssim$8-9 pc at the assumed distance modulus
of 33.3 mag.

Note that we have not discovered a UCD in the bright luminosity regime
$-13.5 \lesssim M_V \lesssim -12.2$, within which also only very few
UCDs are found in Fornax and Virgo.  This confirms the rareness of
these extreme objects. Given our overall survey completeness of
$\sim$50\% and assuming a Poisson distribution for UCD number counts,
we can exclude at 95\% confidence the existence of more than two UCDs
with $M_V<-12.2$ mag and $r_{\rm eff}<70$ pc, within 120 kpc of NGC
4696.

In spite of the lack of such very bright UCDs, there is now a total of
eight confirmed UCDs with $M_V<-12$ mag in Centaurus (this paper,
Mieske et al.~\cite{Mieske07}), comparable to the numbers in
Virgo/Fornax (Jones et al.~\cite{Jones06}, Firth et
al.~\cite{Firth07}). Seven of those eight sources belong to the main
cluster Cen30. Given our completeness of $\sim$50\% within the central
120 kpc of Cen30, we can constrain the true number of UCDs with
$M_V<-12$ mag in that area to 14 $\pm$ 5. How does this compare to the
number of GCs expected from extrapolating a Gaussian globular cluster
luminosity function to $M_V<-12$ mag? For the central Cen30 galaxy NGC
4696, we would expect a total of $\sim$14000 GCs, adopting a specific
frequency of 7.3 (Mieske et al.~\cite{Mieske05}) and an absolute
magnitude of $M_V=-23.2$ mag (Misgeld et al.~\cite{Misgel09}) for NGC
4696. Since we observe within a radius of 120 kpc, we assume that we
include most of the GC systems (e.g., Rhode \& Zepf~\cite{Rhode01}
show the GC system of NGC 4472, the most luminous early-type galaxy in
Virgo, extends to $\sim$80 kpc). Adopting an absolute turnover
magnitude of $M_V=-7.4$ mag (Kundu \& Whitmore~\cite{Kundu01}) and a
GCLF width of 1.35 mag (Jord\'{a}n et al.~~\cite{Jordan06},
\cite{Jordan07}), the expected number of GCs with $M_V<-12$ mag is
about 5. Although being on the low side, this is still within
2$\sigma$ of the estimated number of UCDs. Note that only for
$M_V<-12.8$ mag the expected number of GCs drops below 0.5. From a
purely statistical point of view, only the very brightest UCD
luminosities ($M_V\simeq -13.5$ mag) are thus unaccounted for by a
Gaussian GCLF.

\begin{figure}[h!]
\begin{center}
\epsfig{figure=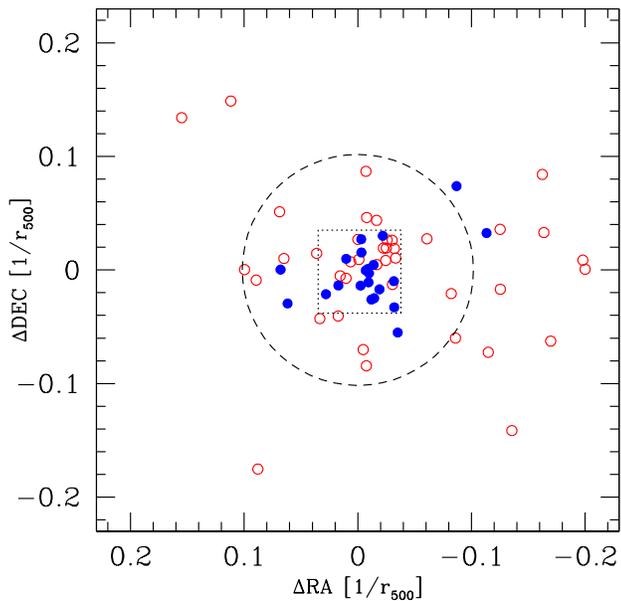,width=8.6cm}
\caption{Positions of confirmed UCDs in the Fornax (open red circles;
Mieske et al.~\cite{Mieske08c}) and Centaurus clusters (filled blue
circles; this paper and Mieske et al. 2007), relative to the central
galaxies and normalised to the $r_{500}$ radius of the respective
clusters. UCDs are selected as compact stellar systems with $M_V<-11$
mag.  The dashed circle indicates the coverage of the search for
bright UCDs in Centaurus presented in this paper. The dotted square
indicates the central quadrant of the P76 Centaurus survey (Mieske et
al.~\cite{Mieske07}), within which the survey completeness was twice
as large as outside (see also Fig.~\ref{map2}). }
\label{mapcenfnx}
\end{center}
\end{figure}

With the database of Centaurus UCDs at hand, it is worthwhile to
compare their spatial distribution to those in Fornax. In
Fig.~\ref{mapcenfnx} we show the projected distance of these two UCD
populations (with UCDs defined as compact stellar systems with
$M_V<-11$ mag) relative to the central galaxies of Centaurus and
Fornax. The Fornax UCD database is the same as used in Mieske et
al.~(\cite{Mieske08c}).  The distances are normalised to the r$_{500}$
radii of either cluster, for which Reiprich \& Boehringer
(\cite{Reipri02}) give r$_{500}$=840 kpc for Fornax, and
r$_{500}$=1.14 Mpc for Centaurus. Within the radius of 120 kpc
surveyed for this publication, the Centaurus UCD population is
slightly more clustered than the Fornax UCD population, at the 96.5\%
confidence level according to a KS-test. However, this may at least
partially be due to the fact that all Centaurus UCDs with $M_V>-12$
mag were discovered in our P76 survey, which had a more complete area
coverage for $r<50$kpc than for $r>50$kpc (see Fig.~\ref{map}). When
restricting to $r<50$ kpc in Centaurus and the corresponding $r<35$
kpc Fornax, the cumulative radial distribution of both samples is
indistinguishable. When considering only UCDs with $M_V<-12$ mag --
for which the spatial survey coverage is comparable between Fornax and
Centaurus -- we also find indistinguishable distributions according to
a KS-test. We can therefore state that within 2$\sigma$, Fornax and
Centaurus UCDs have the same radial distribution, when scaled to the
respective $r_{500}$ radius of their host clusters.

\vspace{0.2cm}

\noindent There are a number of ongoing investigations of the UCD
luminosity function towards very bright luminosities ($M_V<-12$ mag)
in a range of environments, (e.g. the present work, Wehner \&
Harris~\cite{Wehner08}, Misgeld et al.~\cite{Misgel08}). It will be
interesting to investigate how the luminosities/masses of the most
massive UCDs correlate with the properties of their host environments
(Hilker et al. 2009, in preparation), as previously studied in an
analogous fashion for globular cluster systems
(Whitmore~\cite{Whitmo03}, Larsen~\cite{Larsen02}) and systems of
young massive clusters (e.g. Weidner et al.~\cite{Weidne04}).

\label{conclusions}

\begin{table*}
\caption{Properties of the 3 massive UCDs detected in our survey,
ordered by magnitude. Errors are given in parentheses. ``CCOS'' in the
object identifier stands for Centaurus Compact Object Survey, see also
Mieske et al. (~\cite{Mieske07}). The last column gives a 2$\sigma$ upper limit for the half-light
radius in pc estimated from HST WFPC2 archival imaging. }
\label{table}
\begin{tabular}{l|rrrrrrrr}
ID & RA (J2000) & DEC (J2000)&V$_0$ & (V-R)$_0$ & (V-I)$_{\rm 0,estimated}$&  v$_{\rm rad}$ [km/s]& R &r$_h$ [pc]\\\hline\hline
 CCOS J192.129-41.401 & 12:48:30.88 & -41:24:04.96 & 21.15 &  0.51 &  0.77 &   2650 (47) & 8.06 & \\
CCOS J192.200-41.334 & 12:48:48.01 & -41:20:01.52 & 21.22 &  0.64 &  1.03 &   2061 (76) & 5.79 & $\lesssim 8-9$\\ 
  CCOS J191.958-41.258 & 12:47:49.93 & -41:15:28.04 & 21.34 &  0.90 &  1.52 &   3185 (60) & 5.94 & \\ 
 \hline
\end{tabular}
\end{table*}

\end{document}